\documentclass[final,5p,times,twocolumn]{elsarticle}
\usepackage{amsmath}
\usepackage{amssymb}
\usepackage[usenames, dvipsnames]{color}
\usepackage{dcolumn}
\usepackage{bm}
\usepackage{placeins}

\usepackage{listings}
\usepackage{color}
\usepackage[normalem]{ulem}
\usepackage{xspace}
\usepackage{url}
\usepackage{hyperref}
\hypersetup{
    colorlinks,
   citecolor=black,
   filecolor=black,
    linkcolor=blue,
    urlcolor=blue
}

\definecolor{dkgreen}{rgb}{0,0.6,0}
\definecolor{gray}{rgb}{0.5,0.5,0.5}
\definecolor{mauve}{rgb}{0.58,0,0.82}

\newcommand{\sdchannels}{$^3$S$_1-^3$D$_1$\xspace}

\newcommand*{\ket}[1]{\left|{#1}\right\rangle}
\newcommand*{\bra}[1]{\left\langle {#1} \right|}

\newcommand*{\braket}[2]{\left\langle  \left. {#1} \right| {#2} \right\rangle}

\newcommand*{\ketbra}[2]{\ket{#1} \! \bra{#2}}

\newcommand{\editaddblue}[1]{{\color{blue} #1}}

\newcommand{\commentout}[1]{}
\newcommand{\ocite}[1]{}

\begin{document}

\begin{frontmatter}




\title{Nuclear Physics without High-Momentum Potentials: \\
Constructing the Nuclear Effective Interaction Directly from Scattering Observables}


\author[address1,address2]{K. S. McElvain}
\ead{kenmcelvain@berkeley.edu}
  \author[address1,address2]{W. C. Haxton}
\ead{haxton@berkeley.edu}
\address[address1]{Department of Physics,  University of California, Berkeley CA, USA}
\address[address2]{Lawrence Berkeley National Laboratory, Berkeley CA, USA}


\date{\today}

\begin{abstract}
The traditional approach to nuclear physics encodes phase shift information in a nucleon-nucleon (NN) potential,
producing a
nucleon-level interaction that captures the sub-GeV consequences of QCD.
A further reduction to the nuclear scale is needed to produce an effective interaction for soft Hilbert spaces, 
such as those employed in the shell model.  Here we describe an alternative
construction of this effective interaction, from QCD directly to the nuclear scale, that is direct and precise.
This eliminates the need for constructing and renormalizing the high-momentum NN potential.
Instead, continuum phase shifts and mixing angles are used directly at the nuclear scale.  The method
exploits the analytic continuity in energy of HOBET (Harmonic-Oscillator-Based Effective Theory) to connect bound states
to continuum solutions at specific energies.   The procedure is systematic, cutoff independent, and
convergent, yielding keV accuracy at NNLO or N$^3$LO, depending on the channel.  Lepage plots are provided.

\end{abstract}

\begin{keyword}
effective theory \sep nucleon-nucleon interaction \sep phase shifts



\end{keyword}

\end{frontmatter}


The traditional approach to nuclear physics employs an NN potential to encode experimental phase shift information,
which is then renormalized to produce an effective interaction appropriate for soft, discrete bases,
such as those used in the shell model (SM).  
Most often NN potentials (see 
\ocite{CDBonn,Av18,Nijmeigen} \cite{Machleidt:2000ge,  Wiringa:1994wb,Pieper:2001mp,  DeSwart2018}) are determined empirically.  
For example, the Argonne $v_{18}$ interaction \ocite{Av18} \cite{Wiringa:1994wb,Pieper:2001mp}
contains 18 operator components and 40 parameters, adjusted to reproduce pp and np 
scattering data over the energy range 0-350 MeV, as well as low energy nn scattering parameters and the deuteron binding 
energy.  Phenomenological forms are assumed for the associated short- and mid-ranged radial forms, including
correlation functions that build in hard cores at $r \sim 0.5$ fm.

Chiral effective field theory (EFT) provides an alternative to such
phenomenologically derived NN potentials (see the reviews \ocite{Machleidt2011,Epelbaum2009} \cite{machleidt2011chiral,Epelbaum:2008ga} and references therein).
As a systematic expansion, chiral EFT provides a basis for error estimation and for the systematic
inclusion of three- and other multi-nucleon interactions \ocite{Epelbaum3}\cite{Epelbaum:2002vt}, including the order at which these become
important in a given counting scheme.
The starting point is an effective Lagrangian with pion and nucleon fields, and often including an explicit delta. Physics above the 
break-down scale $\Lambda_b \sim 1 \mathrm{~GeV} \sim
m_\rho$ is represented through short-range effective operators, with good convergence expected for momenta $q$
where $\frac{q}{\Lambda_b} << 1$.  Typically such potentials are regulated, as otherwise the introduction of
counterterms to guarantee the convergence of loops becomes tedious.  

Both approaches encode experimental scattering data in an NN potential, which then must be renormalized
to produce an interaction appropriate for SM-like soft spaces.   This two-step renormalization
procedure -- the introduction of a GeV-scale nucleon-level effective theory (ET) in the guise of an  NN potential, then integrating out most of the high momentum content of that potential
in forming a soft nuclear effective interaction -- is somewhat unusual.
In other EFT contexts the reduction is done in one step, from the initial ultraviolet (UV) theory directly to the desired
$P$ (or included) space (here the SM space).  One identifies the general form 
of the effective interaction in $P$ based on a relevant operator expansion, then determines the coefficients of these 
operators (the low-energy constants or LECs) by matching to observables.

The nuclear two-step procedure is tricky to execute well, in part because the natural bases for describing the NN interaction
and multi-nucleon bound states have different properties.   As NN interactions are determined from phase shifts, the natural basis consists of 
continuum plane-wave states.  In contrast, the only discrete and compact Hilbert space available for describing translationally
invariant bound states is the harmonic oscillator (HO):
$P$ spaces containing a complete set of HO Slater
determinants of energy $E \le \Lambda_{SM} \hbar \omega$ (relative to the naive ground state) can be exactly separated into center-of-mass (CM) and relative
motion. Separability of the Hilbert space, while perhaps not an overriding concern in models, is crucial in an
EFT, as it leads to a translationally invariant effective interaction, a great simplification.  The resulting lack of orthgonality  between
NN (plane wave) and nuclear (HO) bases limit the extent to which the NN potential can be softened by methods
discussed below, forcing one to deal with at least a semi-hard core in the subsequent nuclear effective interactions step. 

Early attempts to solve the effective interactions problem quantitatively were typically diagrammatic: the nuclear reaction matrix $G$
was approximated by perturbing in the bare
two-nucleon reaction matrix $G_0$ \ocite{Bethe} \cite{Bethe:1956zz}, generating intermediate particle-hole excitations \ocite{KB} \cite{Kuo1966}.  
This approach was found to fail in the early 1970s.  
Barrett and Kirson \ocite{BK} \cite{Barrett1970}, working in a SM basis, evaluated the effective interaction for $^{18}$F, 
finding large third-order contributions to $G$ that tended to cancel against second-order contributions.
At about the same time Shucan and Weidenmuller \ocite{SW} \cite{Schucan1972, Schucan1973} identified the presence of intruder states -- states of the 
full Hamiltonian that appear within the spectrum of $P$, but reside primarily outside of $P$ -- as a generic source of such nonperturbative
behavior.  

In another early approach, phenomenological super-soft potentials \ocite{TRS} \cite{Haftel:1971er, Cote:1976rlk} were sought in order to make the
nuclear renormalization step more tractable.  
This idea has modern but more systematic analogs in which
a high-momentum NN potential is softened, while not losing physics important to $P$.
A modest reduction of a potential's cutoff scale $\Lambda$ can have great impact
on the numerical complexity of the effective interactions problem:  
the number of single-particle states should diminish as $\Lambda^3$, while the the dimension of the $A$-body Hilbert space 
depends combinatorially on the single-particle basis size \ocite{BognerINT} \cite{Bogner2011}.  Two procedures widely applied
are the $V_{\mathrm{low}~k}$ \ocite{VLK1,VLK2,VLK3} \cite{Nogga:2004ab, Bogner:2006tw,Bogner:2005sn,  Bogner:2005fn,Bogner:2006ai} and the similarity renormalization group (SRG)\ocite{GW,Wegner,BognerSRG1,BognerSRG2} \cite{Glazek:1994qc,Glazek:1993rc, Bogner:2006pc,Jurgenson:2007td}.  
In the SRG approach a continuous sequence of unitary transformations $\hat{U}(s)$ are applied to
the Hamiltonian, indexed by a continuous \editaddblue{flow} parameter $s$, with the variation in $s$ generating a flow equation which can be
exploited to decouple the excluded space from the included space $P$.  The procedure has been carried out in free space
but also with respect to an in-medium reference state \ocite{inmedium} \cite{Hergert:2015awm}.
In the $V_{\mathrm{low}~k}$ approach the $T$ matrix for a potential $V_{NN}^{\Lambda_\infty}$ characterized by a high momentum
cutoff $\Lambda_\infty$ is matched
to one for a low-momentum potential $V_{\mathrm{low}~k}^\Lambda$ characterized by a lower cutoff $\Lambda$.
The matching is done, preserving either the full off-shell or the half-on-shell $T$ matrix, producing 
energy-dependent Hermitian or energy-independent non-Hermitian interactions, respectively.  With certain approximations a
Hermitian energy-independent potential can be obtained.  Both procedures have the
attractive property of integrating out the more model-dependent, short-range behavior of potentials, putting them
into a nearly universal form.

These methods have been used with good success to lower cutoffs scales to $\Lambda \sim 2$ fm$^{-1} \sim$ 400 MeV $\sim$ 2 fm$^{-1}$.  Problems arise if $\Lambda$ is reduced further, cutting into momentum scales important to typical SM \ocite{nocore,Bigstick} \cite{Navratil:2000gs,Navratil:2009ut, Johnson2018} and coupled cluster  and coupled cluster \ocite{CC1,CC2} \cite{Dean:2003vc, Hagen:2013nca} $P$ spaces: one must retain in the softened potential all Fourier components that are numerically significant within $P$.  
As these procedures are typically executed at the NN level,
another issue is the omission of three- and higher-body corrections that grow in importance as $\Lambda$ is lowered.  
In coupled cluster calculations using $V_{\mathrm{low}~k}^\Lambda$,
significant variations in the ground state energies of $^{16}$O, $^{15}$O, and $^{15}$N with cutoff,
$1.6 \le \Lambda \le 2.2$ fm$^{-1}$, have been interpreted as indicating the importance of omitted three-body terms \ocite{Jensen} \cite{Jensen:2010vj}.

Such softened potentials greatly help, but still leave a challenging renormalization step to reach $P$.
Here we describe a direct, one-step procedure for calculating the effective interaction appropriate
for nuclear calculations in translationally invariant HO $P$ spaces.
The scattering data that normally are encoded in a high-momentum potential
are instead used to determine the effective interaction's LECs. The only cutoffs or regulators in the treatment are those defined by $P$ itself (the
oscillator parameter $b$ and $\Lambda_{SM}$).  While the choice of $b$ and $\Lambda_{SM}$ can affect the rate of convergence,
converged results are independent of whatever choice is made.  The theory, which comes in pionless and pionful forms, is highly
convergent, with keV accuracy achieved in NNLO or N$^3$LO, depending on the channel.


HOBET \ocite{HOBETSong,HOBETLuu,HOBETthreebody,HOBETform} \cite{haxton2000morphing, Haxton2001,Haxton2002, Luu2004, Haxton2008},  
uses the energy-dependent
Bloch-Horowitz (BH) equation \ocite{BH} \cite{Bloch1958} to generate the effective interaction within a finite HO $P$ space.  The BH equation
produces exact eigenvalues and exact restrictions of the true wave functions $\Psi$ to $P$.  This attractive definition of effective
wave functions as restrictions to $P$ is a consequence of the energy dependence: as projection onto $P$ does not preserve scalar products,
this property cannot be achieved with Hermitian energy-independent effective Hamiltonians.  BH solutions evolve simply with
increases in $\Lambda_{SM}$, with new components added but old components unchanged.
Another attractive consequence of the energy dependence is the generation of every eigenstate having
 a nonzero overlap with $P$ (in general, an infinite number) even though the Hilbert space $P$ is finite.
Consequently there are no intruder states in BH treatments: any state not 
generated in $P$ does not couple to $P$.  

Despite these attractive properties of BH solutions, there is some prejudice against energy-dependent approaches in nuclear 
physics \ocite{BognerINT,Jennings} \cite{Bogner2011, Jennings2005}.  As was originally demonstrated by Brandow \ocite{Brandow} \cite{BRANDOW:1967zz} via folded diagrams,
the energy dependence can be removed to yield
an energy-independent, non-Hermitian Hamiltonian that preserves the attractive properties of BH equation solutions.
But more commonly a form of Lee-Suzuki \ocite{LS} \cite{Lee:1980yrt, Suzuki1980} transformation is employed to produce an energy-independent
Hermitian interaction reproducing certain eigenvalues, but not the other properties described above.

\begin{figure}[ht]   
\centering
\includegraphics[scale=0.275 ]{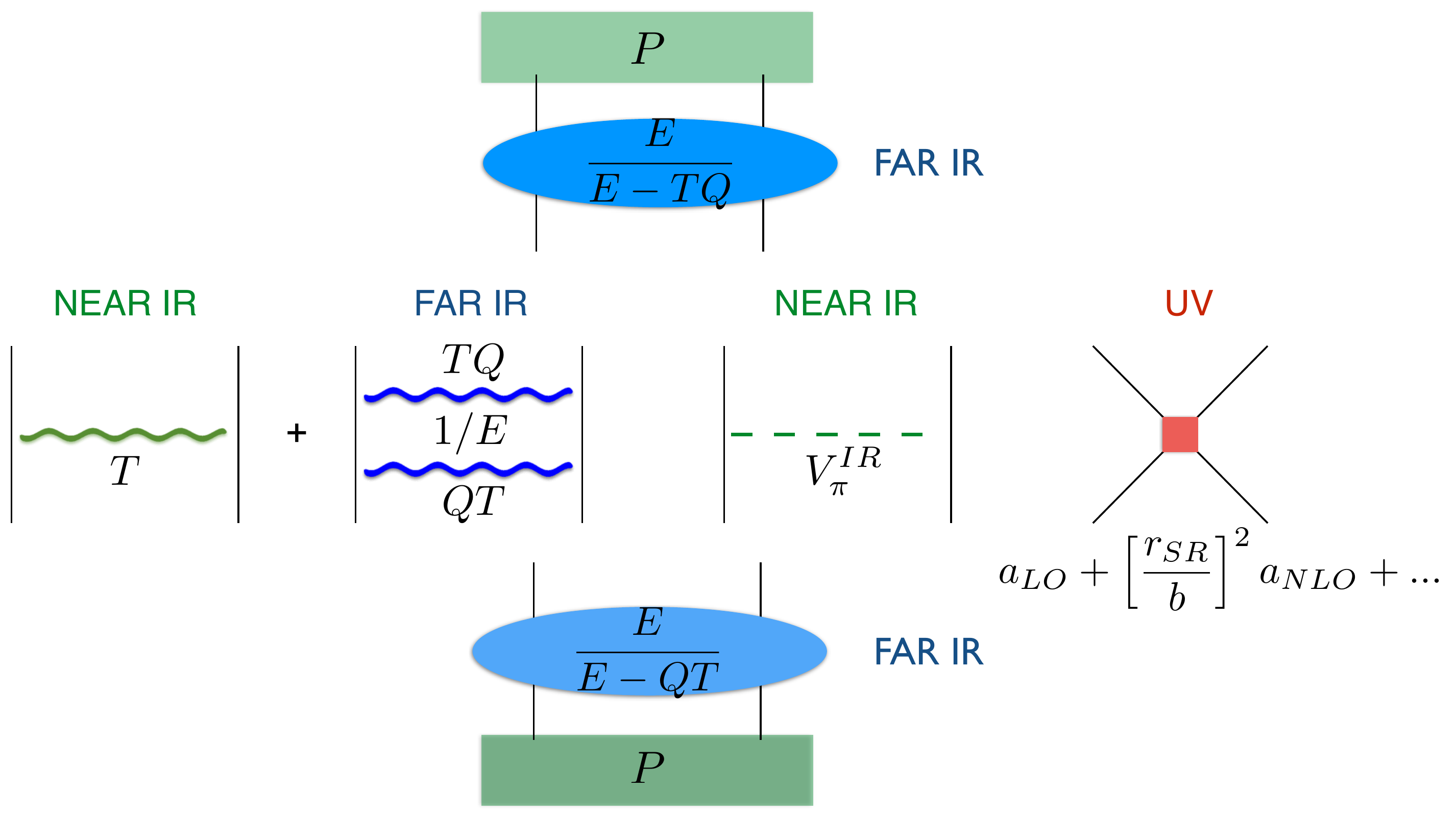}
\caption{HOBET's pionful effective interaction, appropriate to a HO where translational invariance 
requires $P$ to be defined in terms of total quanta (in contrast to chiral interactions employing a momentum regulator). (Color online: blue, green, red
indicate far-IR, near-IR, and UV corrections.)}
\label{Fig_Heff}
\end{figure}

From the perspective of ET, energy-dependent  formulations have another attractive property, preserving 
analytic continuity in energy important to describing both bound and continuum states seamlessly. 
Most of the information carried by NN phase shifts is encoded in their evolution with energy, which is especially rapid near
threshold.  In an ET that keeps energy an an explicit parameter, this information can be used directly and simply.

There are reasons one can offer for eschewing energy-dependent approaches.
One is the need to find self-consistent solutions of the BH equation:
the eigenvalue sought appears as a parameter in the effective interaction used.
But in HOBET calculations performed to date \ocite{HOBETSong,HOBETLuu,HOBETthreebody,
HOBETform} \cite{haxton2000morphing, Haxton2001,Haxton2002, Luu2004, Haxton2008}, energy self-consistency to machine accuracy is achieved very quickly by iteration, typically in 5-6 steps.
It is possible to organize the algorithm so that steps after the first require little work.  Second and perhaps
more serious is the assumption that the LECs of an energy-dependent formulation must themselves depend
on energy -- one envisions a BH effective interaction
resembling Argonne $v_{18}$, except that a distinct set of 40 parameters would be needed at every energy.
Such complexity could easily convince one to seek a different approach.

Key to resolving the second issue is the observation that an ET formulated in a finite HO basis requires corrections
in both the UV and infrared (IR), the former because the hard core is unresolved in $P$, the 
latter because the HO over-confines weakly bound nuclear states.   UV corrections are associated with hard 
short-range scattering that kicks nucleons high into the excluded space $Q=1-P$:  differences in energies
of the initial states in $P$ are of little consequence.  In contrast, IR corrections are governed by the relative
kinetic energy operator $T$, a ladder operator in the HO that couples the last included shell in $P$ with
the first excluded shell of the same parity in $Q$: the lack of any scale separation in the IR leads to
sharp energy dependence and also very slow convergence of IR corrections in perturbation theory \ocite{HOBETLuu} \cite{Haxton2001,Haxton2002}.
The IR is responsible for $\sim$ 95\% of the nuclear BH equation's energy dependence.

These observations motivated the following reorganization of the BH equation to separate IR and UV corrections \ocite{HOBETLuu} \cite{Haxton2001,Haxton2002}. 
\begin{equation} \label{eq:BH}
\begin{gathered}
\quad\quad\quad\quad\quad\quad P H^\mathrm{eff} P\ket{ \Psi } = E P \ket{\Psi},  \hfill\\
G_{QT} \equiv  \frac{1}{E - QT},~~~G_{QH} \equiv \frac{1}{E-QH},~~~H \equiv T+V,  \hfill\\
H^\mathrm{eff} = E G_{TQ}(E)  \left[ T + {T\frac{Q}{E}T} + V + V_\delta \right] E G_{QT}(E), \hfill \\
\quad\quad\quad\quad\quad\quad V G_{QH} QV  \leftrightarrow  V_\delta, \hfill \\
\end{gathered}
\end{equation}
This reorganization yields Green's functions in $QT$ that carry almost all of the energy dependence.
In \ocite{HOBETform} \cite{Haxton2008} is was shown that $V_\delta$ can be readily represented by a contact-gradient expansion with
with constant LECs.  Any residual energy dependence not captured by the Green's functions is easily absorbed by the associate operators.
Consequently $H^\mathrm{eff}(E)$'s LEC parameterization is as simple as that of standard NN potentials.
In \ocite{HOBETform} \cite{Haxton2008} those LECs were determined from the Argonne $v_{18}$ by numerical renormalization.
Here we will show they can be determined directly, without the use of an NN potential, from the variation of phase shifts with energy.

Numerical solutions of Eq. (\ref{eq:BH}) for two- and three-nucleon systems can be obtained for
bound or continuum states in formulations that begin with a potential $V$ \ocite{HOBETLuu,HOBETform} \cite{Haxton2001, Haxton2002, Haxton2008}.
Such solutions exhibit all of the BH equation properties described above.
We discuss the three contributions to Eq. (\ref{eq:BH}), $V_\delta$, $V$, and the kinetic energy, in turn.

HOBET's short-range expansion for $V_\delta$ \ocite{HOBETform} \cite{Haxton2008} is built on the HO creation operators
 $(a^\dagger_x,a^\dagger_y,a^\dagger_z) \equiv a_i^\dagger$  and their conjugates
 \begin{equation} \nonumber
 a_i^\dagger \equiv \frac{1}{\sqrt{2}} \left( -\frac{\partial}{\partial r_i} + r_i \right), \;\quad a_i\equiv \frac{1}{\sqrt{2}} \left( \frac{\partial}{\partial r_i} + r_i \right), \vspace{1pt}
 \end{equation}
which satisfy the usual commutation relations.
Here $\boldsymbol{r}=(\boldsymbol{r}_1-\boldsymbol{r}_2)/\sqrt{2} b$ is the dimensionless Jacobi coordinate.
Defining projections with good angular momentum,
 $a^\dagger_M = \hat{\boldsymbol{e}}_M \cdot \boldsymbol{a}^\dagger$ and $\tilde{a}_M=(-1)^{M+1} a_{-M}$, where
 $\hat{\boldsymbol{e}}_M$ is the spherical unit vector, we can form the scalar HO nodal
 raising/lowering operators $\hat{A}^\dagger \equiv {\bf a}^\dagger \odot {\bf a}^\dagger$,  $\hat{A} \equiv {\bf \tilde{a}} \odot {\bf \tilde{a}}$
 \begin{equation}
  \hat{A}~ \ket{n \ell m } = -2\; \sqrt{\left(n-1\right)\left(n+\ell-{1/ 2}\right) }  \; \ket{n-1 \, \ell m }, \nonumber 
 \end{equation}
where $\ket{n \ell m}$ is a normalized HO state.  Using
\begin{align}
 \delta(\boldsymbol{r}) & = \sum_{n^\prime n} d_{n^\prime n}\commentout{^{00}} \ketbra{ n^\prime 0 0 }{ n 00 }, \hfill  \nonumber \\
 d_{n^\prime n} & \equiv \frac{2}{\pi^2} \left[ \frac{\Gamma(n^\prime +{\frac{1}{2} }) \Gamma(n+\frac{1}{2}) }{ (n^\prime-1)! \, (n-1)!} \right]^{1/2} , \hfill 
 \end{align}
 HOBET's short-range expansion can be carried out, which we note below  is a Talmi moment expansion about the momentum scale $b^{-1}$ \ocite{HOBETform}\cite{Haxton2008}.
 We obtain for the $^1S_0$ channel N$^3$LO interaction
\begin{equation}
\begin{gathered}
V_\delta^{\mathrm{S}} =\sum_{n^\prime n} d_{n^\prime n}\commentout{^{\, 0 \, 0}} \, \Big[ a^{\mathrm{S}}_{LO} \, \ketbra{ n^\prime \, 0 }{ n \, 0}  \hfill \\
\quad\quad\quad + a_{NLO}^{\mathrm{S}} \, \left\{ \hat{A}^\dagger  \ketbra{ n^\prime \, 0 }{ n \, 0 }  +  \ketbra{ n^\prime \, 0 }{ n \, 0} \hat{A} \right\} \hfill \\
\quad\quad\quad+ a_{NNLO}^{\mathrm{S} , 22} \, \hat{A}^\dagger  \ketbra{ n^\prime \, 0 }{ n \, 0 } \hat{A}  \hfill \\
\quad\quad\quad+ a_{NNLO}^{\mathrm{S} ,40} \, \left\{ (\hat{A}^{\dagger \, 2} \ketbra{n^\prime \, 0 }{ n \, 0} +  \ketbra{n^\prime \, 0 }{n \, 0} \hat{A}^2 \right\} \hfill \\
\quad\quad\quad +  a_{N^3LO}^{\mathrm{S} ,42} \, \left\{ \hat{A}^{\dagger \, 2} \ketbra{n^\prime \, 0}{n \, 0} \hat{A} + \hat{A}^\dagger \ketbra{n^\prime \, 0 }{ n \, 0} \hat{A}^2 \right\}  \hfill \\
\quad\quad\quad +   a_{N^3LO}^{\mathrm{S} ,60} \, \left\{ \hat{A}^{\dagger \, 3} \ketbra{n^\prime \, 0}{n \, 0} + \ketbra{n^\prime \, 0 }{ n \, 0} \hat{A}^3 \right\} \Big] , \hfill
\end{gathered}
\end{equation}
 where the LECs $a_{LO}, a_{NLO}, ...$ carry units of energy.  The HO matrix elements are
 \begin{equation} \label{eq:deltaVMe}
 \begin{gathered}
 \langle n^\prime (\ell^\prime=0\,S)JM  | V_\delta^\mathrm{S} \ket{ n(\ell=0\,S)JM } \quad =  \quad d_{n^\prime n}~\biggl[ \hfill \\
\quad\quad a_{LO}^{\mathrm{S}} -2 \bigl[(n^\prime{-}1)+(n{-}1) \bigr] a_{NLO}^{\mathrm{S}} +4 (n^\prime{-}1)(n{-}1)a_{NNLO}^{\mathrm{S},22} \hfill  \\
\quad\quad  +\,4 ((n^\prime{-}1)(n^\prime{-}2)+(n{-}1)(n{-}2)) a_{NNLO}^{\mathrm{S},40} \hfill \\
\quad\quad -8 ((n^\prime{-}1)(n^\prime{-}2)(n{-}1)+ (n^\prime{-}1)(n{-}1)(n{-}2))a_{N3LO}^{\mathrm{S},42} \hfill  \\ 
\quad\quad +\,8((n^\prime{-}1)(n^\prime{-}2)(n^\prime{-}3)+(n{-}1)(n{-}2)(n{-}3))a_{N3LO}^{\mathrm{S},60} \biggr] \hfill
\end{gathered}
\end{equation}

In tensor channels, such as \sdchannels the angular momentum raising and lowering operators are needed, formed
from the fully aligned coupling of the spherical creation and annihilation operators
\begin{equation}
\begin{gathered}
\langle n \ell || \left[ {\bf {a}}^\dagger  \otimes \cdots \otimes{\bf {a}}^\dagger \right]_\ell  || n 0 \rangle = \langle n 0 || \left[ {\bf \tilde{a}} \otimes  \cdots \otimes{\bf \tilde{a}}\right]_\ell  || n \ell \rangle = \hfill  \\
~~~~~~~~~~~~~~  2^{\ell/2}  \sqrt{ \frac{l! }{ (2 \ell-1)!!} \frac{\Gamma[n+\ell+ \frac{1 }{ 2}] }{ \Gamma[n+ \frac{1 }{ 2}]}} , \hfill
\end{gathered}
\end{equation}
where $||$ denotes a reduced matrix element.  By applying the angular momentum raising operator to the delta function expansion one can form operators such as
\begin{eqnarray}
V_\delta^{\mathrm{SD}} &=& \sum_{n^\prime n} d_{n^\prime n}\commentout{^{\,0 \,0}} \, \Big[ a_{NLO}^{\mathrm{SD}} \left\{  \left[  {\bf a}^\dagger \otimes {\bf a}^\dagger \right]_2 \, | n^\prime \, 0 \rangle \langle n \, 0 |  + \right.  ~~~ \nonumber \\
&& \left. | n^\prime \, 0 \rangle \langle n \, 0 | \,  \left[  {\bf \tilde{a}} \otimes {\bf \tilde{a}} \right]_2 \right\}\odot \left[ \boldsymbol{\sigma}_1 \otimes \boldsymbol{\sigma}_2 \right]_2  + \cdots \Big]  .
\end{eqnarray}
Full results through N$^3$LO for all contributing channels can be found in \ocite{HOBETform} \cite{Haxton2008}.

Equation (\ref{eq:deltaVMe}) shows that HOBET's  ladder operator expansion generates a characteristic dependence on nodal quantum
numbers $n,n^\prime$: $a^S_{LO}$ is the only LEC contributing to the HO $1s$-$1s$ ($n$=$n^\prime$=1) matrix element, $a^S_{NLO}$ is the
only additional LEC contributing to the $1s$-$2s$ matrix element, etc.   Consequently if one starts with an NN potential --
the two-step process described previously for either a hard potential like Argonne $v_{18}$ or a softer one like
$V_{\mathrm{low}~k}$ -- the LECs can be fixed in a scheme-independent way, once one computes 
individual matrix elements of the effective interaction.  If $a^S_{LO}$ is determined from the $1s$-$1s$ matrix element in a LO calculation, that
value will not change at NLO, and so on.  The $s$-wave LECs in this scheme are proportional to 
\begin{equation} \label{eq:TalmiInt}
 \int r^{\prime \, 2} dr^\prime r^2 dr  ~ r^{\prime \, 2 (n^\prime-1)} e^{-{r^\prime}^2/2} \,
 V_Q(r^\prime, r) \, r^{2( n-1)} e^{-{r}^2/2},
 \end{equation}
 with $V_Q \sim V G_{QH}QV$: this identifies the LECs as a 
 nonlocal generalization of Talmi integrals \ocite{Talmi} \cite{DeShalit1963}.

$V_\delta$ impacts the interpretation of the term linear in the potential, written as $V$ in  Eq. (\ref{eq:BH}).
 If we are given a potential $V$, its short-range contributions will enter in the low-order
 Talmi integrals, for which there are LECs to fix the values, up to the order of the expansion.  That is, one can decompose 
 $V$ into UV and IR components, $V=V^{UV}+V^{IR}$, with $V^{UV}$ denoting the part of $V$ 
 contributing to Talmi
 integrals where LECs are available, and $V^{IR}$ the remainder.  Only $V^{IR}$ is relevant:  in any fit to observables, 
 the effects of
 $V^{UV}$ can be absorbed into the  LECs of $V_\delta$.  
 
 There are three natural choices for $V$.  In the two-step treatment where $V$ is given, one can treat it as is,
 knowing only the long-range part $V^{IR}$ will matter.  Alternatively, we can sever all connections to $V$,
 building a true ET in $P$, 
 following one of two paths:  1) a pionful ET, with $V \rightarrow V_\pi^{IR}$, building in the correct long-distance NN behavior;
 or 2) a pionless ET, with $V \rightarrow 0$.
 
 In other EFT approaches pion exchange is frequently treated as an interaction between point nucleons,
 producing a $1/r^3$ tensor force that must be regulated.  That is, the cost of building in the proper 
 long-distance behavior of the NN interaction through an explicit pion is the introduction of a short-range contribution that is both poorly behaved
 and unrealistic, as the nuclear potential is dominated at short distance by vector mesons, not the pion.
 In HOBET $V$ is naturally regulated by its embedding in $P$, and as noted above, operationally plays
 no role at short range, where LECs are available.  For example,
 in an N$^3$LO calculation the leading-order s-wave contribution of $V_\pi$ -- the first Gaussian moment
 not fixed by an available LEC -- is
  \[ \int r^2 dr \,  r^8 e^{-r^2} V_\pi(r) \sim V_\pi^{IR}. \]
The integrand peaks at $|\vec{r}_1-\vec{r}_2| \sim 4.1$ fm (taking $b$=1.7 fm), far out on the tail of the pion exchange
potential.  Consequently, in the depiction of pionful HOBET of Fig. \ref{Fig_Heff}, the term in
Eq. (\ref{eq:BH}) linear in $V$ has been replaced with $V_\pi^{IR}$ and labeled as a near-infrared contribution.

\begin{figure*}[ht]   
\centering
\includegraphics[scale=0.51]{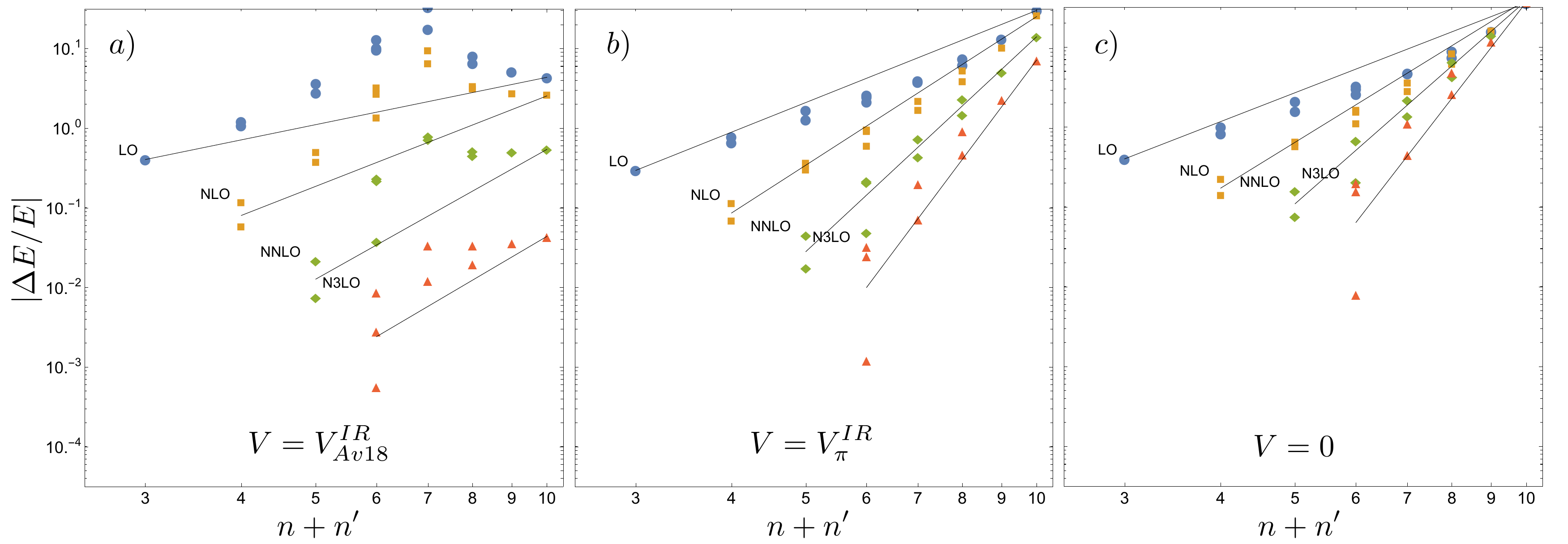}
\caption{The Lepage plots for scheme-independent fitting to the s-wave matrix elements of $H^{eff}$
in the deuteron ($^3S_1-{}^3D_1$) channel for the Argonne $v_{18}$ potential.  The fractional error in $V^{eff}$ is plotted vs.
the sum of the nodal quantum numbers.  The panels correspond 
to Eq. (\ref{eq:BH}) with a) $V=V^{IR}
_{A v_{18}}$; b) $V=V_\pi^{IR}$ (pionful HOBET), and
c) $V=0$ (pionless HOBET).  See text.}
\label{fig:Lepage}
\end{figure*}
 
 Lepage plots for these three cases -- $V^{IR}$ equated to $V^{IR}_{Av18}$, $V^{IR}_\pi$, and 0 for Argonne $v_{18}$, pionful HOBET,
 and pionless HOBET calculations, respectively --  are given in Figure \ref{fig:Lepage}, where the fractional error $|\Delta E/E|$
 in matrix elements of $V^{eff} \equiv E \, G_{TQ}(E)(V+V_\delta) E \, G_{QT}(E)$ are plotted as a 
 function of the sum of the nodal quantum numbers $n+n^\prime$.  These are evaluated for the deuteron $^3S_1$-${}^3D_1$ bound state at -2.2246 MeV,
 using $b$=1.7 fm and $\Lambda_{SM}=8$.
 We use the scheme-independent fitting procedure described previously, as that choice cleanly divides
 the low $n,n^\prime$ matrix elements used in fitting LECs from those of higher $n,n^\prime$, which are predictions.
 Only the latter are plotted.
 The straight lines in the figure are drawn from the fractional error at maximum $n+n^\prime=10$ to that at 
 minimum $n+n^\prime$ (averaged over the possible values).  
The steepening of the trajectories with
 increasing order demonstrates that the improvement is systematic.  While the convergence is all three cases is quite 
 satisfactory, the use of scheme-independent fitting in this comparison unduly favors the potential treatment:  the proper way to
 fit the LECs in pionful and pionless HOBET is described below.  The steeper trajectories for pionful HOBET shows the 
 advantages of building in our knowledge of the NN interaction's pion tail.
 As discussed in \ocite{HOBETform} \cite{Haxton2008}, the order-by-order convergence 
of the short-range expansion $V_\delta$, apparent from Fig. \ref{fig:Lepage}, is governed by the implicit dimensionless parameter $(r_{SR}/b)^2$, where 
$r_{SR}$ represents the range of the unresolved short-range physics.  
In pionful HOBET one would expect $r_{SR}$  to be determined by vector meson or effective sigma masses \cite{oshima2015};
in pionless HOBET $r_{SR}$ would be the typical range of the strong interaction.   

The Green's functions in $V^{eff}$ alter matrix elements only in cases where $n$ or $n^\prime$ resides in
the last included shell of $P$, immediately below $Q$.  All other components of $P|\Psi \rangle$ are annihilated by $QT$, so that
$\frac{E}{E-QT} \rightarrow 1$.  We
caution that $\frac{E}{ E-QT} P | \Psi \rangle$ should not be misconstrued as attaching an IR  ``tail" to the wave function -- that is,
as something akin to a  Woods-Saxon \ocite{WS} \cite{Woods:1954zz} or J-matrix \ocite{Jmatrix} \cite{Alhaidari2008} modification of a HO state. 
Rather, the Green's functions are a component of the effective interaction, part of the BH $H^{eff}$.
The $P$ space continues to be the compact HO space described by $b$ and $\Lambda_{SM}$ -- a special space due to its
separability.

The remaining terms in Fig. \ref{Fig_Heff}, which depend only on $T$, correct for the effects of HO over-confinement on the
kinetic energy.   They can be rearranged to form a rescattering series
\begin{equation} 
 P \left[ T + T \, Q \, \frac{T}{E}  + T \, Q  \, \frac{T}{E} \, Q  \, \frac{T}{E} + \cdots \right] P=P T \frac{E }{ E-QT} P,
\end{equation}
and summed (see below).
The terms generated from $QT$ account for the delocalization that occurs in weakly bound physical states.  The shift (relative
to the simple HO estimate) grows to 
  $-\hbar \omega$, for a bound state just below threshold.
 
 The edge state can be computed from the free Green's function, at the cost of a matrix inversion in $P$
 \begin{eqnarray}
 E G_{QT}P |n \, \ell  \, m \rangle &=& G_0(E) [P G_0(E) P]^{-1} \ket{n \, \ell \, m}, \nonumber \\
 G_0(E) &=& \left\{ \begin{array}{ll}  1/(\boldsymbol{\nabla}^2 - \kappa^2) & E<0 \\ 1/(\boldsymbol{\nabla}^2 + k^2) & E>0\, .\end{array} \right. ~~~
 \label{eq:freeG}
 \end{eqnarray}
A homogeneous term $\phi(E)$ can be added on the right, a freedom we will exploit to build in the correct boundary conditions
for our continuum states.    Here  $\kappa \equiv \sqrt{2 |E|/\hbar \omega}$, $k=\sqrt{2E/\hbar \omega}$, and $\boldsymbol{\nabla}$ are
dimensionless.  Matrix elements of $G_0$ in $P$ can be evaluated analytically.  
We employ standing-wave Green's functions.  

The proper treatment of these Green's functions is important to our main goal, a consistent one-step procedure for
determining HOBET's $H^\mathrm{eff}$ directly from scattering data, rather than through the two-step process
of constructing then renormalizing a potential.

In the case of bound states, as described in earlier work \ocite{HOBETform} \cite{Haxton2008}, 
self-consistent solutions of the BH equation are obtained only at the eigenvalues $E$, for a given choice of LECs.
$G_0(E)$ depends only on $E$.  Thus if an eigenvalue $E$ is known -- the simplest example is the deuteron bound
state -- one should demand a solution at that $E$.  This becomes an implicit constraint on the LECs.
Just as the deuteron binding energy is used in parameterizations of conventional potentials,
the LEC $a_{LO}^{{}^3\!S_1}$ can be determined by demanding a BH solution at $E=-2.2246$  MeV.  

However, most of our information on the NN interaction comes from phase shifts and mixing angles, and thus from continuum states.
HOBET treats bound and continuum states on an equal footing, in each case generating the restrictions of the full wave functions to $P$.
But unlike the bound-state state case, in general there exists a solution at every energy $E>0$.  The self-consistency constraint now
comes from the fact that the Green's function depends not only on $E$, but also the phase shift $\delta_\ell(E)$.  The experimental
phase shift is used directly in the nuclear $H^\mathrm{eff}$, inserted through the homogeneous term in the kinetic energy Green's function, rather than in
a nucleon-level potential,
\begin{equation}
\begin{gathered}
G^\ell_0(E>0,\delta_\ell(E) ;{\bf r}, {\bf r}^\prime) =\displaystyle{ - \frac{\cos{k|{\bf r} - {\bf r}^\prime|} }{ 4 \pi |{\bf r}-{\bf r}^\prime|} }  \hfill \\
\quad\quad - k  \cot{\delta_\ell(E)}~ j_\ell(k { r}) ~j_\ell(k { r}^\prime)~\sum_{ m}Y_{\ell m}(\Omega) Y_{\ell m}^*(\Omega^\prime). \hfill
\end{gathered}
\end{equation}
When the resulting $H^\mathrm{eff}$ is diagonalized, in general an eigenvalue at the selected $E$ will not be found.
As the theory is complete and the IR behavior correct, the source of this discrepancy must
be in the UV, an inadequate $V_\delta$.  $V_\delta$'s  LECs should then be adjusted to fix the discrepancy.  

LECs are chosen to produce a best fit to all of the phase shift information from threshold to a ``fuzzy" maximum
in the CM energy, through the procedure described below.  The relevant experimental information depends on the order of the HOBET expansion
and the choice of $P$ space.  For  N$^3$LO and the $P$ used in our study ($\Lambda_\mathrm{SM}=8$, $b$=1.7 fm),
the relevant data correspond to CM energies $\lesssim$ 50 MeV.  
The channels that enter at N$^3$LO 
are  $^1$S$_0$, \sdchannels, $^1$D$_2$, $^3$D$_1$, $^3$D$_2$, $^3$D$_3$-$^3$G$_3$,
$^1$P$_1$, $^3$P$_0$, $^3$P$_1$, $^3$P$_2$-$^3$F$_2$, $^1$F$_3$, $^3$F$_3$ and $^3$F$_4$.
The number of LECs at N$^3$LO varies from six in the S-wave channels to one in the F-wave and mixed DG-wave channels. 
In these fits  $V_\pi$ with pion mass dependence is taken from Eq. 17, 18 and 19 of \ocite{Av18}\cite{Wiringa:1994wb} using the recommended coupling constant value $f^2=0.075$.  Smaller intermediate range contributions from Eq. 20 in the same paper corresponding to two pion exchange have been omitted.  The regulator $(1 - e^{-cr^2})$ has also been removed as the potential is automatically regulated by the $P$-space basis.
As a crosscheck on this procedure, N$^3$LO LEC fits were also done in the $^1\!F_3$ channel with phase shifts at 2, 5, 10, 15, 20, 25, 30, 40, and 50 MeV in which ${f}^2$ was treated as a
second LEC, together with the N$^3$LO LEC.  The fit yielded a very similar value $f^2=0.74$, demonstrating numerically that pion exchange dominates the NN potential
at the long distances where $V_\pi^{IR}$ contributes.

\begin{figure}[b]   
\centering
\includegraphics[scale=0.335 ]{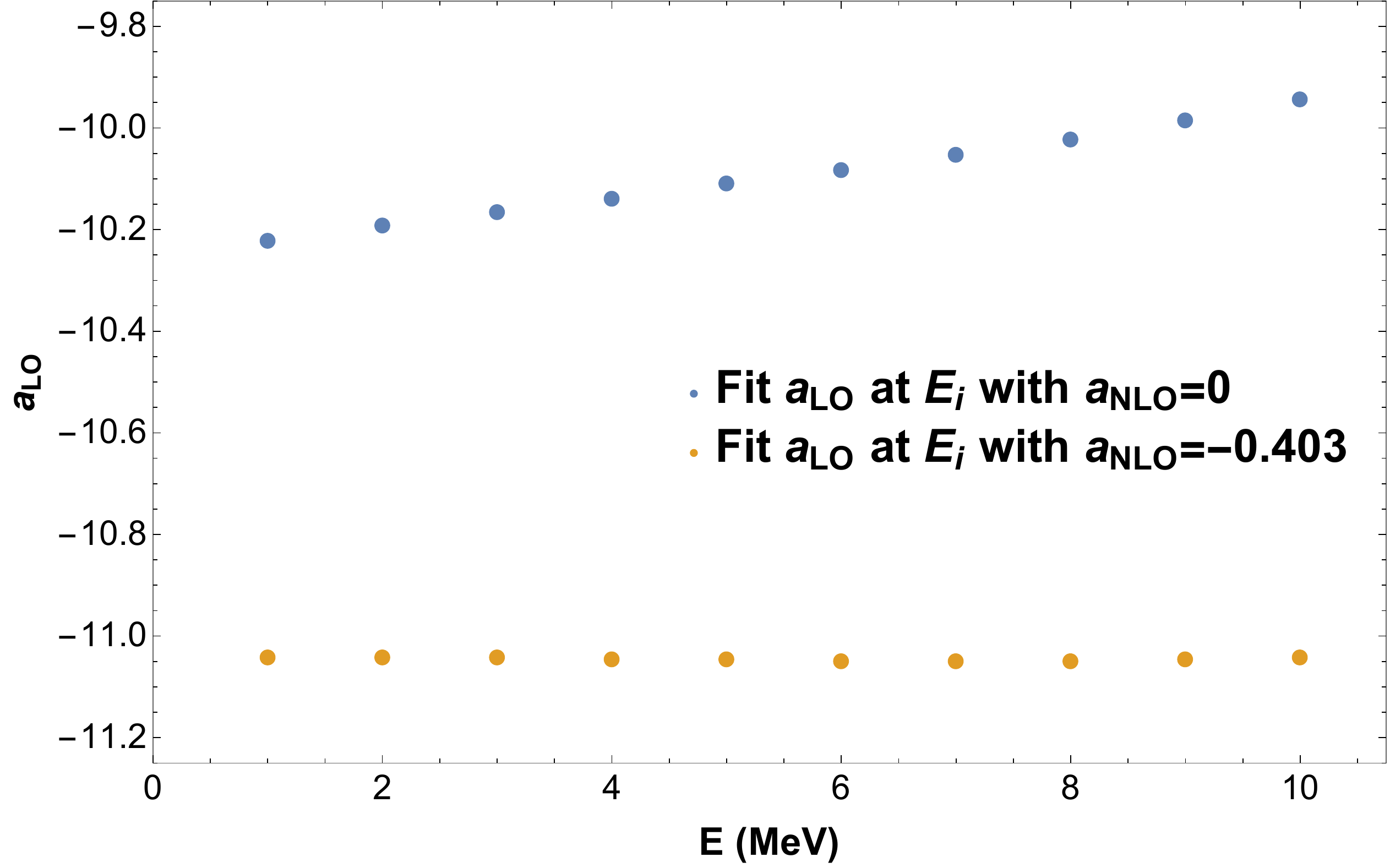}
\caption{Energy dependence of $a_{LO}$ at LO (upper dots) and residual energy dependence $a_{LO}$ at NLO (lower dots) after $a_{NLO}$ is fixed at -0.403.     }
\label{Fig:LECEnergyDep}
\end{figure}
Before we tackled the fitting of LECs with experimental phase shifts, we performed a numerical experiment with an analytic S-wave model -- 
a square well plus hard core resembling the nuclear potential --  for which exact scattering parameters can be derived.  This
experiment influenced the procedures we designed.
We solved for $H^{eff}$ using a HOBET $P$ space with $b=1.7$ fm and $\Lambda=8$ (5 included S-states).  Rapid convergence was found, a $\sim$ two-orders-of-magnitude improvement in $\chi^2$ per order 
in the expansion \ocite{HM2} \cite{McElvainPrep}. 
We then did a series of calculations to explore the consequences of omitted higher-order operators on the LECs of retained, lower-order operators.  
First, we worked through a set of 10 energies $E_i$ equally space from 1 to 10 MeV, using Green's function with the appropriate model phase shifts at the $E_i$,
solving for $a_{LO}$ by requiring  $H^{eff}(E_i)$ to yield $E_i$.
The ten values we obtained are shown as the upper blue dots in Fig. \ref{Fig:LECEnergyDep}: a slight energy dependence in the determined $a_{LO}$ is apparent, about 3\% over the energy range.  
Second, we then repeated the fit at two energies $E_i$, 1 and 10 MeV, but at NNLO, adjusting $a_{LO}$ and $a_{NLO}$ so that again the $H^{eff}(E_i)$ yielded $E_i$.
Third, keeping $a_{NLO}$ fixed at the value determined in the second step, we  repeated the initial set of LO calculations.  
The resulting $a_{LO}$s, shown as the lower gold dots in Fig. \ref{Fig:LECEnergyDep},
now exhibit almost no dependence on energy ($< 0.1$\%). 

This and other experiments provided strong evidence the average effect of omitted higher-order operators is to a very good approximation absorbed by the operators of the last included order.
This allows us to simplify the fitting of LECs in high-order calculatiuon, through a bootstrap or iterative  procedure.  
 For example, for a fit at the NNLO, the LO LECs can be taken from the previous NLO fit and held fixed; only the NLO and NNLO LECs need to be fit, with previously determined 
 values for the NLO LECs serving as reasonable initial values.   In a subsequent N$^3$LO fit, the LO and NLO LECs would be kept fixed at the values determined in the NNLO
 calculation, with only the NNLO and N$^3$LO terms adjusted.
 
 We then applied this procedure to realistic NN scattering.  The phase shifts and mixing angles we use in testing our procedure are those generated from Argonne $v_{18}$.  
Because the potential's parameters are carefully fit to scattering data, numerically these phase shifts can be regarded as experimental ones.
But unlike experiment data with errors, this gives us a potential and a set of scattering data that are precisely equivalent at each energy, which is helpful for some the tests we describe below.
 
Several interesting issues that arise in the fitting procedure are described in detail elsewhere \ocite{Kthesis} \cite{McElvain2017}, and thus are treated briefly here.  
Given a potential and a consistent set of scattering data, one can compare the scheme-independent LEC fitting
 procedure of \ocite{HOBETform}\cite{Haxton2008} with the new procedure described above.   In the earlier procedure, individual matrix elements of $H^{eff}$ were calculated numerically from Argonne $v_{18}$,
 with  $a_{LO}$ then determined from the $1s$-$1s$ matrix element, etc. We found that the $H^{eff}$ determined with the current scheme does a significantly better job in representing scattering data \ocite{Kthesis}\cite{McElvain2017} than the earlier scheme-independent method.  We attribute this to the ability of the new fit, through the last included order,
 to absorb the average effects of omitted higher-order operators.

The fitting is done at selected continuum energies (or equivalently momenta); in the case of the \sdchannels  channel, one can 
also choose to use the bound state.   What grid of points should be used in fits?
The resolution of unity in the  channel $\ket{\ell,m}$ is
\begin{equation}
\begin{gathered}
 1 = \sum_{i \in \mathrm{bound}} |i \, \ell \, m \rangle \langle i  \, \ell \, m| + \frac{2}{\pi} \, \int_0^\infty dk \ketbra{k \, \ell \, m}{k \, \ell \, m} \hfill \\
\braket{ r\; }{ k \, \ell \, m } \equiv k r \left[ -\cos{\delta_\ell} j_\ell(k r) +\sin{\delta_\ell} \eta_\ell(k r) \right]\, . \hfill
\label{eq:resolution}
\end{gathered}
\end{equation}
As the integral is weighted in $dk$, not in $dE$, we select points evenly spread in $k$. ( Note that with experimental data or phase shifts obtained from lattice QCD calculations,
we would not have this freedom, and thus a more sophisticated weighting of points might be needed.)

The number of sample points must at a minimum exceed the number of LECs to be fit: in practice considerably more are used.
The adequacy of the continuum grid selected can be checked by increasing the density of points to verify that consistent LECs are obtained.

What range of continuum momenta should be used in the fit, and how should grid points be weighted?  Generally the LECs of an ET 
for some low-energy $P$ are determined from the longest wavelength information available.  If the ET is well behaved,
 then once its LECs are determined, other long-wavelength observables can be predicted, including those somewhat beyond the 
 momentum or energy scale
 used in the LEC fitting.  Intuitively one anticipates that a LO theory would utilize very long wavelength information in its LEC fitting, and be valid
 only over a limited range of energies or momenta.  Additional input at somewhat shorter wavelengths would be need to determine the LECs
 of a NLO theory, and the resulting ET would be valid over a somewhat more extended range, and so on.  Our LEC fitting
 procedure is designed to emphasize data from an energy range appropriate to the order of the fit being done.

This is accomplished through a cost function that takes into account the potential impact of operators beyond the order 
being considered.  Fits are done over a set of energies substantially larger than the number of LECs being determined.
With an exact ET and perfect phase shift data, $PH^\mathrm{eff}(E_i) P|\Psi_i \rangle = E_i P |\Psi_i \rangle$
for each energy $E_i$ of a set spanning the energy interval of interest.  But as the ET is only executed to some specified order $N$,
$PH^\mathrm{eff}(E_i) P|\Psi_i \rangle = \epsilon^N_i P |\Psi_i \rangle$, where $\epsilon^N_i$ is an eigenvalue
near but not identical to $E_i$.  We  determine the LECs by minimizing the cost function 
\begin{equation}
\chi^2_{\mathrm{order}~N} = \sum_{i \in \{ \mathrm{sample} \} } \frac{(\epsilon^N_i - E_i)^2 }{ \sigma_{N+1}(i)^2 } ,
\end{equation}
where $\{ \mathrm{sample} \}$ represents the set of energy points used, in the case of unmixed channels such as $^1S_0$ and $^3P_0$.

The variance $\sigma_i^2$ is an estimate of the contributions of omitted higher-order LECs not included
in the fit,
\begin{equation}
\sigma^2_{N+1}(i) \sim  \kappa^2_{N+1} \sum_{ \{ a_j^{N+1} \} } \left(  \frac{ \partial \epsilon_i^{N+1} }{ \partial a_j^{N+1} } \biggr\rvert_{a_j^{N+1} =0} \right)^2 .
\label{eq:var}
\end{equation}
Here $\epsilon_i^{N+1}$ is the eigenvalue at one order beyond that being employed in the fit, and
$\{ a_j^{N+1} \}$ is the set of LECs that contribute in that order.  Under the assumption that the values of these LECs are uncorrelated,
the change in the energy $(\epsilon_i^N-E_i)^2$ that would result from turning on the $\{a_j^{N+1} \}$ can be estimated from the sum over the squares of first-derivative variations
in each of the directions $a_j^{N+1}$, evaluated at $a_J^{N+1}=0$.   These would then be folded with an estimate of the typical scale 
of such variations, represented by  $\kappa^2_{N+1}$ in Eq. (\ref{eq:var}), which in a direct calculation at order $N+1$ would be computable
from the values obtained for the LECs for order ${N+1}$.  Absent such a calculation, $\kappa^2_{N+1}$ can be estimated from the lower-order LECs,
under the assumption of naturalness.   In the present treatment, the value of $\kappa^2_{N+1}$ is irrelevant, as it acts a common
scale factor in $\chi^2_{\mathrm{order}~N}$, and thus does not alter the relative weightings of energy points in our sample.  {(This would
not be the case were we fitting experimental phase shifts with errors, as Eq. (\ref{eq:var}) would then include a second term reflecting 
those errors -- an uncertainty in the energy to which one should assign an experimental phase shift.)}

In the LEC fittng, the $\sigma^2_{N+1}(i)$ generate a soft cutoff in the energies over which we sample.
In  a LO calculation, a high weight is placed on energy points where the NLO contribution is expected to be small,
and a low weight on those where the NLO contribution is large:  $\sigma_i^2$ increases as the energy $E_i$ is raised.  
The net effect of the resulting cost function is to limit the range of contributing
phase shifts to low energies.  The range grows with increasing order, reflecting the greater importance of higher energy scattering data
to higher order LECs.  

Numerically it proved helpful to perform fits successively, e.g., with the final results for the NLO LECs used as the starting values in the search for the best
values of N$^2$LO LECs, and so on.  This improves the rate of convergence in higher orders, where the cost function
minimization is over multiple LECs \ocite{Kthesis} \cite{McElvain2017}.

The above discussion applies to single channels:  in mixed channels, such as \sdchannels, one obtains for a given energy $E_i$
two standing-wave solutions, which due to the typically small mixing value $\Sigma$ will be mostly S channel and mostly D channel,
\begin{eqnarray}
|\Psi_S \rangle &=& \cos{\Sigma} |S(\delta_S) \rangle - \sin{\Sigma} |D(\delta_S) \rangle, \nonumber \\
|\Psi_D \rangle &=& \sin{\Sigma} |S(\delta_D) \rangle + \cos{\Sigma} |D(\delta_D) \rangle,
\label{eq:basis}
\end{eqnarray}
with the indicated phase shifts, where the notation corresponds to the Blatt-Biedenharn \ocite{BB} \cite{Blatt:1952zza} parameterization of the S-matrix
\begin{equation}
\!\hat{S} = \hat{O}^{-1}\! \left( \begin{array}{cc} e^{2 i \delta_S} & 0 \\ 0 & e^{2 i \delta_D} \end{array} \right) \! \hat{O},~~\hat{O} = \left( \begin{array}{cc} \!\cos{\Sigma} & {-}\sin{\Sigma} \\
\!\sin{\Sigma}  & \cos{\Sigma} \end{array} \right).\!\!\!
\end{equation}
The general standing wave solution can be written as a mixture of the basis states given in Eq. (\ref{eq:basis}), with
probabilities $\cos^2{\alpha}$ and $\sin^2{\alpha}$ for $|\Psi_S \rangle$ and $|\Psi_D \rangle$, respectively.
The single-channel sampling is generalized for mixed channels by including in the sampling not only grid points in $k$,
but values $\alpha$ of 0, $\frac{\pi}{ 4}$, and $\frac{\pi }{ 2}$ at each $k$.  This allows us to access the three
degrees of freedom in the S-matrix, $\delta_S$, $\delta_D$, and $\Sigma$.  Details are given in \ocite{Kthesis} \cite{McElvain2017}.

We then applied our procedures to our Argonne $v_{18}$-equivalent scattering database, for both 
pionful and pionless HOBET.   
For fitting 41 phase shift data samples are used, evenly spaced in $k$, and running from 1.0 to 80.0 MeV.    80 MeV is well beyond the point where N$^4$LO LECs are needed and demonstrates the effectiveness of the soft cutoff created by $\sigma_{N+1}^2$.
In the coupled channel case if $\cot\delta_D > 100.0$ we drop the sample for numerical reasons in the construction of the Green's function for $G_{QT}$.  By carrying out the fitting program from LO through N$^3$LO, 
we obtained a series of LEC sets defining a progression of HOBET potentials of increasing sophistication.
The \sdchannels  deuteron bound-state energy was not included in the fitting, and therefore becomes a prediction.
Table \ref{tab:deuteron} shows the results as a function or order.

While both pionful and pionless calculations converge well,
the comparison shows the importance of the including the pion,
which we stress again is explicitly an IR correction in HOBET.
At N$^3$LO the deuteron binding energy is correct to 3 keV, and the phase-shift fit (reflected in
the self-consistency error) is nearly perfect. Table \ref{table:1} gives the N$^3$LO \sdchannels and $^1$S$_0$ LECs obtained;
results for other channels can be found in \ocite{HM2} \cite{McElvainPrep}. 

\begin{table}
\caption{\label{tab:SPotEnergyTable} 
Deuteron channel: binding energy $E_b$ as a function of the expansion
order.  Bare denotes a calculation with $T+V_{IR}$ and no IR correction. The error columns are the average of squared fractional error up to 10 MeV.}
\begin{tabular}{lcccc}
\hline\hline
\textrm{Order}&
\textrm{$E_\mathrm{b}^\mathrm{pionless}$}&
Error & \textrm{$E_\mathrm{b}^\mathrm{pionful}$}&
Error \\
\hline
bare & 3.0953 & - & -0.67187 & - \\
LO      &  -0.9214 & 1.16E-2 & -2.0206 & 1.84E-3 \\
NLO   &  -1.5392 & 1.53E-3  & -2.17814 & 3.44E-5 \\
NNLO & -1.6267 & 1.37E-3 & -2.1952 & 3.32E-5  \\
N$^3$LO & -2.0690 & 1.34E-4 & -2.2278 & 6.07E-6 \\
\hline\hline
\end{tabular}
\label{tab:deuteron}
\end{table}

\begin{figure}[htbp]   
\centering
\vspace{2mm}
\includegraphics[scale=0.3 ]{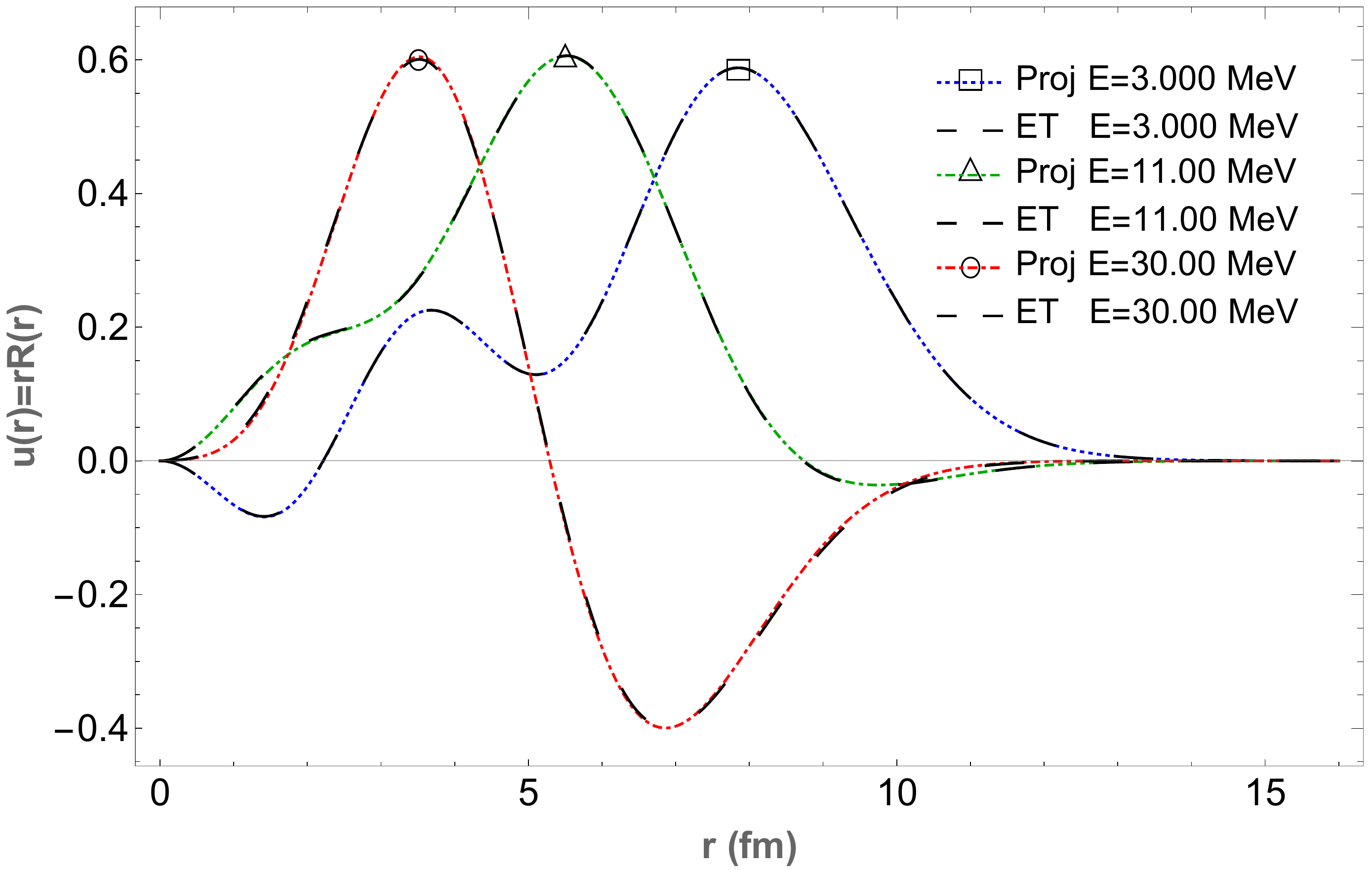}
\caption{Projections of exactly computed $^1P_1$ relative wave functions (colored, dotted lines) are shown to match the HOBET wave
 functions(large black dashes) nearly perfectly, for representative continuum energies. The results are predictions: the selected energies are
 distinct from those used in LEC fitting.}
\label{Fig:Fit1P1}
\end{figure}
We also compared the HOBET  ${}^1P_1$ wave functions to projections of the exactly computed wave functions at energies 3, 11, and 30 MeV, 
which were deliberately chosen to be distinct from the sample energies 2.55, 3.22, 10.14, 11.45, 28.83, and 31.0 MeV, to ensure that results are not directly 
constrained by our fitting.  Fig. \ref{Fig:Fit1P1} shows that the wave functions are in nearly perfect agreement: all of the detailed behavior of the projected wave functions as continuous functions of $r$ and $E$, remarkably, can be encoded in a few energy-independent LECs, provided the leading energy dependence is first treated via the $QT$ summation of Eq. (\ref{eq:BH}).

We contrast this behavior with that found in an alternative procedure \cite{Binder:2015trg}.  
The decomposition of the BH equation
in Eq. (\ref{eq:BH}) with its summation of $QT$ to all orders, was done to cure an IR pathology of the HO,
namely that the hoping operator $T$ induces strong nonperturbative coupling of $P$ and $Q$ involving nearest-neighbor
shells of the same parity.  The kinetic energy resummation removes this coupling, thus restoring the scale separation 
necessary for a well-behaved EFT \cite{Haxton2001,Haxton2002, Luu2004}.
Recently Binder et al. \cite{Binder:2015trg} directly embedded a momentum based EFT, where the kinetic energy is diagonal and thus generates
no coupling of $P$ to $Q$, in a finite HO basis,
taking matrix elements of the interaction operators against the discrete set of kinetic energy eigenstates while preserving the LECs of the EFT.    
The LECs are then adjusted to create a phase shift match at the corresponding momenta via the J-matrix method \cite{Alhaidari2008}.  
In contrast to the behavior illustrated in Fig. \ref{Fig:Fit1P1}, the resulting behavior is not smooth, 
with the predicted phase shifts oscillating between the momentum points used in the fitting.   This is indicative of a forced fit,
not a predictive EFT.

\begin{table}
\caption{The deuteron channel and S-wave LECs determined at N$^3$LO in pionless and pionful HOBET.  See \ocite{HM2} \cite{McElvainPrep} for the full set of couplings.}
\label{table:1}
\begin{tabular}{cccc}
\hline\hline
Transitions & LECs (MeV) & Pionless& Pionful \\
\hline
\rule{0pt}{3ex} ${}^3S_1 \leftrightarrow {}^3S_1$  & $a_{LO}^{3S1} $ & -50.9105 & -47.2779  \\ [1pt]
                                                      & $a_{NLO}^{3S1}$  & -4.3625 & -5.14528 \\ [1pt]
                                                      & $a_{NNLO}^{3S1,22}$ &  1.8670E-2 & -9.6852E-1  \\ [1pt]
                                                      & $a_{NNLO}^{3S1,40}$ &  -2.2203E-1 & -2.4459E-1 \\ [1pt]
                                                      & $a_{N3LO}^{3S1,42}$  &  2.3691E-2 &  -1.3784E-1 \\ [1pt]
                                                      & $a_{N3LO}^{3S1,60}$  & -6.7398E-2 & -4.7928E-2 \\ [1pt]
\rule{0pt}{3ex} ${}^3S_1 \leftrightarrow {}^3D_1$  & $a_{NLO}^{SD} $ & -2.6731 & -9.4681  \\ [1pt]
                                                      & $a_{NNLO}^{SD,22}$  & -6.8852E-1 & -3.0647 \\ [1pt]
                                                      & $a_{NNLO}^{SD,04}$  & 3.4194E-1 & -1.4228 \\ [1pt]
                                                      & $a_{N3LO}^{SD,42}$  & -7.3097E-2 & -4.8398E-1 \\ [1pt]
                                                      & $a_{N3LO}^{SD,24}$  & -2.3028E-2 & -7.3943E-1 \\ [1pt]
                                                      & $a_{N3LO}^{SD,06}$  & 9.1250E-2 & -5.3541E-2 \\ [1pt]
\rule{0pt}{3ex} ${}^3D_1 \leftrightarrow {}^3D_1$  & $a_{NNLO}^{3D1} $ & 4.5685 & 3.2278  \\ [1pt]
                                                      & $a_{N3LO}^{3D1}$  & 8.7938E-1 & 9.1347E-1 \\ [1pt]
\rule{0pt}{3ex} ${}^1S_0 \leftrightarrow {}^1S_0$  & $a_{LO}^{1S0} $ &  -38.5612 &  -38.5364 \\ [1pt]
                                                      & $a_{NLO}^{1S0}$ & -5.7331 & -5.9948  \\ [1pt]
                                                      & $a_{NNLO}^{1S0,22}$ & -8.8427E-1 & -1.2224  \\ [1pt]
                                                      & $a_{NNLO}^{1S0,40}$ & -3.9656E-1 & -4.2192E-1  \\ [1pt]
                                                      & $a_{N3LO}^{1S0,42}$ & -6.5638E-2 & -1.5812E-1  \\ [1pt]
                                                      & $a_{N3LO}^{1S0,60}$ & -3.8120E-2 & -4.1352E-2  \\ [1pt]
\hline\hline
\end{tabular}

\end{table}

In summary, we have demonstrated a precise method to construct the effective interaction needed at the
nuclear scale, directly from experimental phase shifts.  The only regulators that enter in this one-step method are those
defining the soft nuclear Hilbert space $P$ itself, namely $b$ and $\Lambda_{SM}$.  Thus one can avoid the usual procedure in which scattering data are first encoding
in a high-momentum potential, then decoded through a series of potential
softening and renormalization steps, with associated approximations.  The method exploits HOBET's explicit continuity in energy,
which allows one to connect NN scattering information at a specified energy to the properties of a bound state
at a different energy, without approximations.  The pionless and pionful theories both converge at the nuclear momentum scale, with the pionful theory producing a 
N$^3$LO deuteron binding energy accurate to $\sim$ 3 keV.

HOBET generates not only exact eigenvalues (to the tolerance achieved in the
expansion) but also wave functions that correspond to the exact projections of bound or continuum states to $P$.
Such wave functions evolve simply with changes in $P$, e.g., an increase in $\Lambda_{SM}$ simply adds new
components to the wave function, leaving others unchanged.  While HOBET's convergence can be slowed by picking
a non-optimal $P$, observables are independent of this choice, provided the expansion is carried out to the requisite order.
That is, answers are independent of the regulators $b$ and $\Lambda_{SM}$.
These various properties are attractive in an ET.

A precise connection between nuclear properties and scattering data has important implications for relating nonrelativistic nuclear structure to
lattice QCD (LQCD).  Phase shifts calculated from LQCD \ocite{WL} \cite{Berkowitz:2015eaa, Beane:2013br, Murano2014} can be used directly in our HOBET method.  The
matching to LQCD can be done to LQCD eigenvalues computed
in a finite rectangular volume, by confining HOBET to the same volume.  The method,
which exploits the attractive transformation properties of HO wave functions between Cartesian and spherical bases, will
be described elsewhere \ocite{HM2} \cite{McElvainPrep}.

{\it Acknowledgement:} This material is based upon work supported in part by the US DOE, Office of Science,
Office of Nuclear Physics and SciDAC under awards DE-SC00046548,
DE-AC02-05CH11231, KB0301052, and DE-SC0015376.  We thank Christian Drischler for helpful discussions.

\bibliographystyle{elsarticle-num}
\bibliography{hobetplb}

\end{document}